# MERLIon CCS Challenge Evaluation Plan

# Version 1.2

Leibny Paola Garcia Perera[1], YH Victoria Chua[2], Hexin Liu[2], Fei Ting Woon[2], Andy WH Khong[2], Justin Dauwels[3], Sanjeev Khudanpur[1], Suzy J Styles[2]

[1] John Hopkins University, [2] Nanyang Technological University, [3] TU Delft

February 17, 2023

**Changelog**

Version **1.2** (17-02-2023)

- Added amendments to audio segment id naming under Scoring tool and format of results file and Appendix C.
- Added clarifications on evaluated regions for Task 2 (Language Diarization) under Scoring and File formats and distribution.
- Added clarification on what a valid submission means for the compulsory competition track: Task 1 (Language Identification) Closed Track and increased daily submission limit from 1 to 3 under Evaluation Rules.

Version **1.1** (13-02-2023)

- Updated planned schedule
- Added system description format under Appendix D.
- Added descriptions of results output format under Scoring tool and format of results file and Appendix C.
- Added clarifications on reference annotations in development set and description of reference timestamps in evaluation set of Task 1 under File formats and distribution.
- Added clarifications on scoring speech segments with overlapping language labels and the unit of time under Scoring and Task 2: Language Diarization.
- Added description of baseline system under Baseline System.

## Introduction

The inaugural MERLIon CCS challenge focuses on language identification and diarization for non-standard, accented, spontaneous, code-switched, child-directed speech collected via Zoom. The challenge is intended to:

- benchmark current and novel language identification and language diarization systems in a code-switching scenario including extremely short utterances;
- test the robustness of such systems under accented speech;
- encourage the research community to propose novel solutions in terms of adaptation, training, and novel embedding extraction for this particular set of tasks;

Techniques developed in the challenge may benefit other related fields allowing greater understanding of how code-switching occurs in real-life situations. The results of the challenge will be presented at a special session in Interspeech 2023 in Dublin, Ireland. The tasks evaluated in the challenge are (a) language identification, which is the task of determining the language spoken in a given audio segment, marked by pre-established time markers and (b) language diarization, which is the task of determining what language is spoken when with no pre-established time stamps for speech or language. The Development set and Evaluation set are Challenge-ready partitions of a larger dataset from the Talk Together Study (Woon et al, 2021), and will be provided by the organizers.

In both tasks, there is a closed and open track which differ in the limitations placed on the training set:

- *Closed track.* Participants are only allowed to train their systems on a preselected partition of public data. Training data beyond the preselected partition is not allowed.
- *Open track.* Participants are allowed to train their systems with the preselected partition, as well as up to 100 hours of any publicly available or proprietary data. In the open track, publicly available pretrained models are allowed.

Additional guidelines for open and closed tracks are outlined in Training Data section of this evaluation plan. Participation in the challenge is open to all who are interested and are willing to comply with the rules laid out in this evaluation plan. While there is no cost required to participate in the challenge, participants are encouraged to submit a paper to the corresponding Interspeech 2023 special session. Accepted papers will be presented at Interspeech 2023 in Dublin, Ireland in August 2023. The data, scoring software, and baseline systems are provided free of charge.

For further information consult the sections below, check the FAQs on the website ((https://sites.google.com/view/merlion-ccs-challenge), join the MERLIon CCS mailing list or contact us at merlion.challenge@gmail.com.

**Schedule**

- Registrations Open: 18 Jan 2023

- Registrations Close: 18 Feb 2023

- Training Data Partitions Release: 25 Jan 2023

- Data Release (Development Set): 27 Jan 2023

- Baseline System Release: 13 Feb 2023

- Evaluation Data Release: 16 Feb 2023

- Leaderboard active: 16 Feb 2023

- Official Evaluation Closes (Leaderboard Freeze): 28 Feb 2023

- System Description Submission: 28 Feb 2023

- INTERSPEECH Paper Submission Closes: 1 Mar 2023

- INTERSPEECH Paper Update Submission Closes: 8 Mar 2023

- Leaderboard Reopens (optional further development): 10 Mar 2023

- INTERSPEECH Acceptance: 17 May 2023

## Task Definitions

### Task 1: Language Identification

For the language identification task, the goal is to automatically detect and label the language spoken in an utterance given ground-truth timestamps in an audio recording. During development, systems will be provided with audio recordings where ground-truth language labels have been annotated with timestamps. Each audio segment will have a unique language label, either English or Mandarin. During evaluation, audio recordings are provided with timestamps, but no language labels will be given.

As the audio segments are derived from multispeaker recordings, the audio segment could contain overlapping speech or non-speech vocalizations from other speakers such as infant babbling and crying, coughs, sneezes, laughs, or any noise produced by the speaker by means of their vocal apparatus. However, in the event of overlapping speech, the languages of all speakers are the same, i.e., each audio segment only contains one language.

### Task 2: Language Diarization

For the language diarization, in each audio file, the goal is to automatically detect the length of which language is spoken. During development, systems will be provided with audio recordings where ground-truth timestamps and language labels have been annotated. During evaluation, only audio will be provided. As the recordings feature a variety of code-switching practices as well as multiple speakers, the audio file may contain either only one or both of the target languages (English and Mandarin). In the event of overlapping speech:

- where the languages are not the same (e.g., English segment overlaps with Mandarin segment), both speech segments will be considered for evaluation.
- where the languages are the same (e.g., English segment overlaps with another English segment), both speech segments will be considered for evaluation.

Note that in version 1.0 of the evaluation plan, overlapping language segments with different language labels were not considered for evaluation. However, to ensure fair evaluation of conversational code-switching speech, language overlaps of different language labels are considered in the evaluation as well. When a speech segment occurs with another speaker's non-speech vocalizations, only the region which contains the start and end of the evaluated language segment (i.e., English or Mandarin) will be considered for evaluation.

The Development and Evaluation sets are made of discrete audio recordings from the Talk Together Study; there is no recording that appears in both sets. Timestamp information from the evaluation dataset for Task 1 (Language Identification) is unsuitable for Task 2 (Language Diarization).

**Tracks**

There are open and closed tracks for both tasks, thus there will be a total of 4 sub-tracks in this challenge:

- **Task 1 (Language Identification) Closed Track** – using only a preselected partition of training data listed by organizers

- **Task 1 (Language Identification) Open Track** – up to 100 hours of additional data

- **Task 2 (Language Diarization) Closed Track** – using only a preselected partition of training data listed by organizers

- **Task 2 (Language Diarization) Open Track** – up to 100 hours of additional data

Participation in the closed track for Task 1 (Language Identification) is compulsory for all teams participating in the challenge, while participation in any track for Task 2 (Language Diarization) is optional. Refer to the section *Training Data* in the evaluation plan for more explanation as to what training data can be used in the respective tasks.

**Scoring**

**Scoring metrics**

For Task 1 (Language Identification), system output will be scored by comparing to human reference language labels. The primary evaluation metric will be:

- Equal error rate

The equal error rate (EER) corresponds to the threshold at which the false alarm and miss rates are equal. Accuracy in this work denotes the ratio between the number of correctly classified utterances and the total number of test utterances. The false acceptance rate is computed as the number of false positives, divided by the sum of false positives and true negatives. The false rejection rate is computed as the number of false negatives, divided by the sum of true positives and false negatives. In this challenge, a false rejection occurs when the system incorrectly labels an English segment as Mandarin, while false acceptance occurs when the system incorrectly accepts a Mandarin segment as English.

A secondary evaluation metric for Task 1 (Language Identification) will be:

- Balanced accuracy

Balanced accuracy is defined as the average of recall (or sensitivity) obtained on each class. In this challenge, it will be computed by first calculating the number of English segments correctly predicted by the system, divided by the total number of reference English segments in the recording, thus giving a recall rate for English segments. This is repeated for Mandarin segments, yielding a recall rate for Mandarin segments. The balanced accuracy for each file is then computed as the average of recall rates for English and Mandarin segments.

For this task, only the English and Mandarin segments within each recording are scored. Overlapping language segments with different language labels will not be scored. For instance, for a speech segment labelled as English overlapping with another speech segment labelled as Mandarin, both speech segments are tagged as overlapping with a different language label (see *File formats and distribution*) and will be excluded in the scoring. English and Mandarin segments that overlap with non-speech will be scored.

For Task 2 (Language Diarization), system output will be scored by comparing to human reference language labels and timestamps. The primary evaluation metric will be:

- Language diarization error rate

The language diarization error rate (DER) is computed as the sum of:

- Language error – percentage of scored time for which the wrong language id is assigned within a speech region.
- False alarm speech – percentage of scored time for which a nonspeech region is incorrectly marked as speech containing English or Mandarin.
- Missed speech – percentage of scored time for which a speech region containing English or Mandarin is incorrectly marked as not containing speech.

The unit of time is in milliseconds.

The secondary evaluation metric for Task 2 (Language Diarization) will be:

- Individual language error rates

Mandarin language error rate is computed as the percentage of scored time for which a speech region containing Mandarin is incorrectly marked as non-speech or English. English language error rate is computed as the percentage of scored time for which a speech region containing English is incorrectly marked as non-speech or Mandarin. The unit of time is in milliseconds.

For this task, the scoring region is the total of all annotated portions in the recording. The starts and ends of evaluated regions are demarcated and are excluded from the scoring. Overlapping language segments will be included in the scoring for Task 2 (Language Diarization). Systems should consider non-speech as well as other languages apart from English and Mandarin as non-evaluated regions when constructing the segment boundaries. For more information on the language labels in the challenge dataset, please refer to *Segmentation* section below. For each recording, there are some regions where no annotations have been performed, i.e., there may be speech, but it is unlabeled for timestamps as well as language labels. These regions are not evaluated.

**Scoring tool and format of result files**

All scoring will be performed via scoring scripts available at the MERLIon CCS Github repository (https://github.com/MERLIon-Challenge/merlion-ccs-2023).

**Format of result files.** For Task 1 (Language Identification), a single system output file that consists of all language scores for all audio segments in the evaluation set is required. Each audio segment must have the corresponding language scores for English and Mandarin. All audio segments labelled as English and Mandarin and do not overlap with another language must be included. The order of the audio segments must be according to the order it is presented in the evaluation timestamps labels.

The audio segment id for each audio segment is the combination of audio filename and utt_id and the start and end times (in milliseconds) of that segment, separated by an underscore. Example:

<audio filename>_<utt_id>_<start>_<end>

The relevant information of each field is released under the reference annotations for MERLIon CCS Development set and the reference timestamps for MERLIon CCS Evaluation set for Task 1 (Language Identification).

The scoring script for Task 1 (Language Identification) allows for two formats for the system output file. The first format requires each audio segment to have separate English and Mandarin scores per line. In this format, the English score is always denoted by 0 and must always precede the Mandarin score that is denoted by 1. Thus, each audio segment should have 2 lines, where each line has 3 space-delimited fields. For instance:

| | | |
|---|---|---|
| <audio segment id 1> | 0 | <score of language label> |
| <audio segment id 1> | 1 | <score of language label> |
| <audio segment id 2> | 0 | <score of language label> |
| <audio segment id 2> | 1 | <score of language label> |

…

In the second allowed format for Task 1 (Language Identification) results file, each audio segment is accompanied by their English and Mandarin scores in the same line, where each line has 3 space-delimited fields. For example:

| | | |
|---|---|---|
| < audio segment id 1> | <score of English> | <score of Mandarin> |

< audio segment id 2>          <score of English>          <score of Mandarin>

…

For Task 2 (Language Diarization), a separate RTTM file should be generated for each audio recording. In the RTTM file, each line containing three space-delimited fields:

<start time>     <end time>     <language id>

Where start time and end time refers to the onset and offset of language turn, indicated by language id in milliseconds. The RTTM file for each audio file should be named according to the audio filename.

For visual examples of how the results files for both tasks should look like, refer to Appendix C. For additional details about scoring tool usage, please consult the documentation on the GitHub repository.

## Data

### Training data

For the closed tracks of both Tasks 1 and 2, participants are only allowed to use the following datasets to train their systems:

- 100 hours clean speech from LibriSpeech ASR Corpus [https://www.openslr.org/resources/12/train-clean-100.tar.gz]
  - Note: There are other mirrors available[1]. Only the dataset labeled as *training set of 100 hours "clean" speech* (i.e., train-clean-100.tar.gz) should be used.
- 200 hours of preselected partition of Aishell[2] - filenames in preselected partition available on challenge website.
- 100 hours of preselected partition from the Singapore National Speech Corpus[3] (Koh et al., 2019) - provided by organizers; link available on website.
- Mandarin-English Code Switching in South-East Asia (LDC2015S04)[4]

---

[1] The LibriSpeech ASR corpus is available at https://www.openslr.org/12. For the closed tracks of the challenge, only the partition labelled as training set of 100 hours "clean" speech (i.e., train-clean-100.tar.gz) can be used.
[2] THE AISHELL corpus is available at https://openslr.org/33. For the closed tracks of the challenge, only the preselected partition can be used. Filenames of the files within the preselected partition are available at https://bit.ly/merlion-ccs-aishell-filenames.
[3] The Singapore National Speech Corpus is available at https://www.imda.gov.sg/nationalspeechcorpus preselected partition for the Singapore National Speech Corpus. For the closed tracks of the challenge, only the preselected partition can be used. The preselected partition is available at https://bit.ly/merlion-ccs-nsc-partition.
[4] Also known as the SEAME corpus (https://catalog.ldc.upenn.edu/LDC2015S04).

For MERLIon CCS Challenge purposes, registered participants of the challenge are allowed access to Mandarin-English Code Switching in South-East Asia (also known as SEAME). The SEAME corpus will be distributed by the Linguistic Data Consortium (LDC). For more information on how to access the SEAME corpus, please refer to Appendix A or the challenge website – Datasets. To download the pre-selected partitions of National Speech Corpus and Aishell, please refer to the MERLIon CCS challenge website – Datasets. Refer to Appendix A on the procedure of selection for partitions from Aishell and the National Speech Corpus.

For participation in the closed tracks, training data beyond the above preselected partitions is *expressly prohibited*.

For the open tracks of both Task 1 (Language Identification) and Task 2 (Language Diarization), participants are allowed to use the pre-selected partitions listed above as well as an extra maximum 100 hours of any publicly available or proprietary data to train their systems. Pretrained models that are publicly available (e.g., HuBERT, wavLM) are allowed. Proprietary pretrained models are not allowed. Documentation of the exact model and version number must be reported in the system description document. The above-mentioned limit of 100 training hours can be used for finetuning the pretrained models. Failure to document the pretrained model name and the version number accurately will result in an invalid submission.

For a list of suggested training corpora, please refer to Appendix B. All additional training (or finetuning) data should be thoroughly documented in the system description document. The system description document must also state how custom partitions were selected. If random selection or algorithmic selection has been performed, the procedure should be clearly described, e.g., first 100 hours of audio that fulfills a criterion or first 100 hours randomly selected. Failure to document training data accurately will result in an invalid submission.

**Development and Evaluation data sets**

The Development dataset and Evaluation dataset for the MERLIon CCS challenge (Chua et al., 2022) are Challenge-ready subsets of audio recordings from the Talk Together study (Woon et al., 2021). In the Talk Together study, parents narrated an onscreen wordless picturebook to their child over Zoom video-conferencing software. Each recorded session lasted between 3 and 35 minutes, of which 2 to 25 minutes have been manually annotated by a team of multilingual transcribers focused on linguistic research. Each file in the dataset has been crosschecked by at least 1 senior member of the transcription team.

The Challenge dataset contains 305 Zoom audio recordings of 112 parent-child pairs. 103 of the parent-child pairs were recorded at least twice at separate timepoints, with a maximum of three recordings for a pair. The Challenge dataset includes over 25 hours of child-directed speech in English and over 5 hours of child-directed speech in Mandarin Chinese. Almost all adult voices in the dataset are female parents of a child under the age of 5. Voices of male family members, grandparents and children also occur in the dataset.

**Diverse Recording Environments.** Zoom calls were conducted in the homes of participating families, on a variety of internet-enabled personal electronic devices including laptops, tablets and mobile phones. Recordings were far-field using internal microphones. Environmental background noise varied widely during recordings.

**Diverse Accents.** Adults in our dataset use the Singaporean variety of English, which features different pronunciation from Standard US, Standard UK and other well-documented varieties of English, and also features some unique vocabulary and grammar. Adults in our dataset use the Singaporean variety of Mandarin Chinese, which features different pronunciation from the standard variety of Mandarin (Putonghua), and other well-documented Chinese varieties (Lee, 2010), and also features some unique vocabulary and grammar.

**Diverse Language Use.** The Challenge dataset includes frequent code-switching within and between utterances. Only 61 recordings feature one language throughout. For parents who used both languages, the proportion of Mandarin spoken overall ranged from 0.85% to 80.7%. The utterances are short; on average: 1.4 seconds for English and 1.2 seconds for Mandarin. A full breakdown of the composition of files in the Development and Evaluation sets can be found below.

**Data division (Development and Evaluation).** The Challenge Dataset (Chua et al., 2022) is divided into a Development dataset (for system development) and an Evaluation Dataset (for test). The Evaluation Dataset is selected to be a representative subset of the data, where features such as ratio of English to Mandarin per recording are controlled (Figures 2 to 4; Table 1). Ground truth language labels and timestamps as well as timestamps of evaluated regions will be provided for the Development dataset but withheld for the Evaluation dataset. To reduce overfitting to individual parent-child pairs, the individual parent-child pairs in the Evaluation dataset do not appear in the Development dataset. The voices of researchers are, however, present in both Development and Evaluation datasets.

Table 1. Length of recordings, language segments and speaker characteristics in Development and Evaluation sets of the MERLIon CCS challenge data.

| MERLIon Challenge Dataset (from Talk Together) | Dev | Eval |
|---|---|---|
| Total duration of audio recordings | 28:36:28 | 28:47:14 |
| Number of recordings | 150 | 154 |
| Number of parent-child pairs | 56 | 56 |
| Recordings with one language only | 28 | 33 |
| Recordings with more than one language | 123 | 121 |
| Hours of English speech | 16:09:27 | 15:57:02 |
| Hours of Mandarin speech | 03:04:05 | 03:11:04 |
| Hours of other language speech (non-evaluated) | 00:00:14 | 00:00:19 |
| Number of English segments | 40287 | 39473 |
| Number of Mandarin segments | 9983 | 9766 |
| Median length of English segment (ms) | 1125 | 1120 |
| Median length of Mandarin segment (ms) | 900 | 930 |
| Mean length of English segment (ms) | 1443.84 | 1454.73 |
| Mean length of Mandarin segment (ms) | 1106.41 | 1173.87 |
| Mean proportion of Mandarin: English per recording | 0.51 | 0.52 |

In the Development set, systems will be provided with audio recordings where ground-truth timestamps and language labels have been annotated. In addition, for Task 2 (Language Diarization), the timestamps of evaluated regions are released as well (Table 4 under *File Formats and Distribution*).

In the Evaluation set, for Task 1 (Language Identification), audio recordings with timestamps of audio segments to be evaluated will be provided, while for Task 2 (Language Diarization), audio recordings will not have timestamps. The timestamps of evaluated regions will not be released so as to delink the audio information between Task 1 (Language Identification) and Task 2 (Language Diarization). Timestamp information from the Evaluation dataset for Task 1 (Language Identification) is unsuitable for Task 2 (Language Diarization).

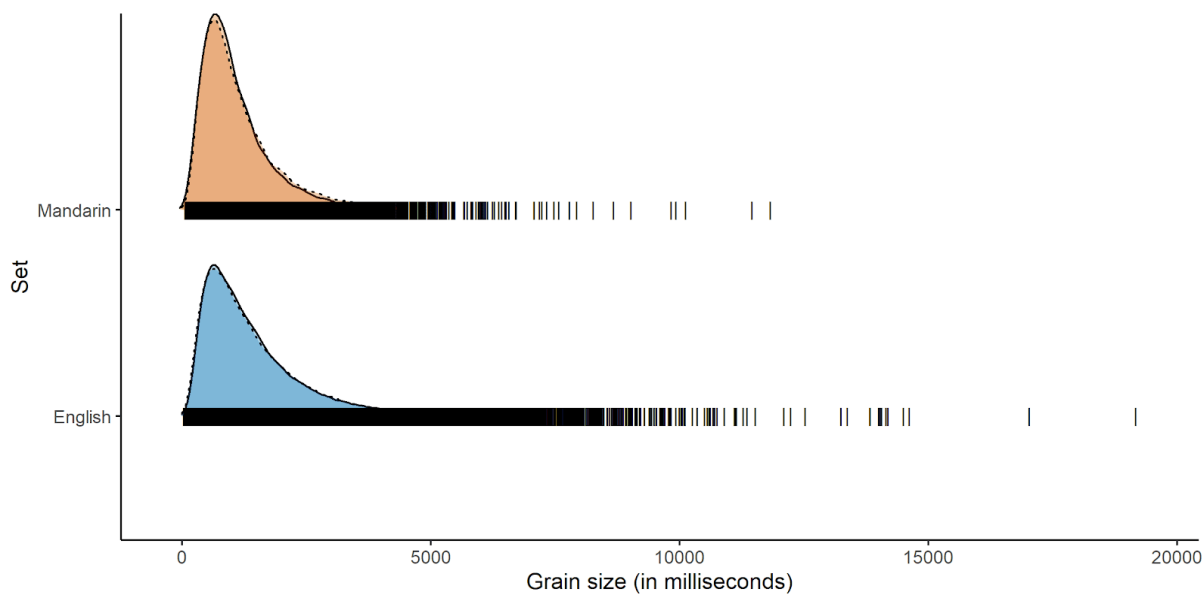

Figure 2. Density plots of the lengths of English and Mandarin grain sizes (in ms) in the Development set (Solid line) and Evaluation set (Dashed line). The strip plots below each density plot indicates the distribution of all grains over both Development set (N = 9983 Mandarin grains; 40287 English grains) and Evaluation set (N = 9766 Mandarin grains; 39473 English grains).

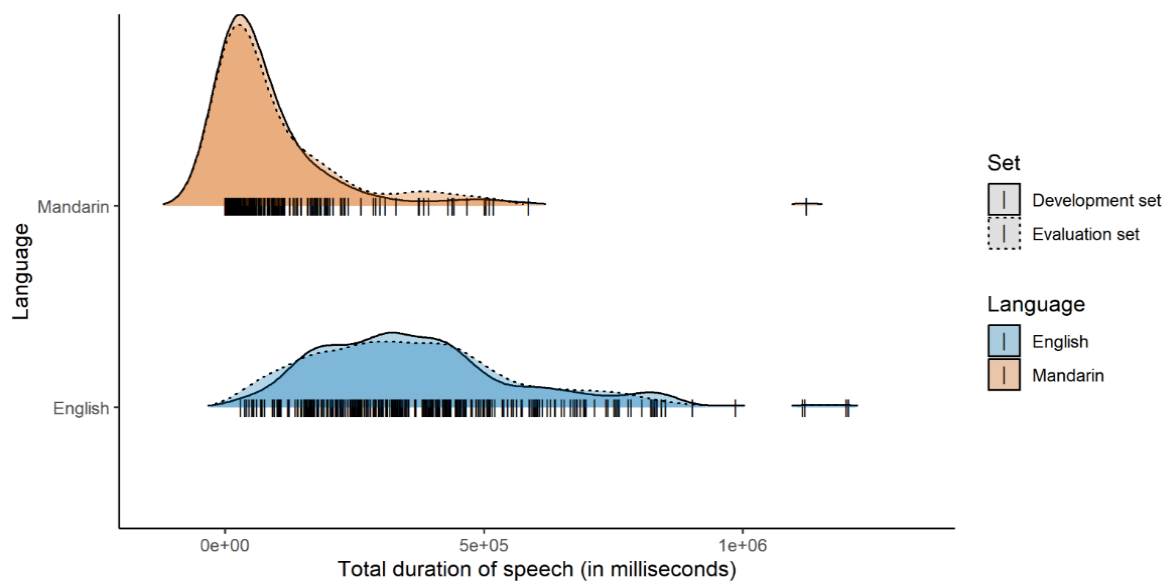

Figure 3. Density plots of the total lengths of English and Mandarin speech (in ms) in each file in the Development set (Solid line) and Evaluation set (Dashed line). The strip plots below each density plot indicate the distribution of lengths in each audio file over both Development set (N = 151) and Evaluation set (N = 154).

**Segmentation.** As part of the Talk Together Study (Woon et al., 2021), each audio recording was manually annotated by transcribers on ELAN, using the in-house BELA transcription protocol (Loh et al., 2022a; https://blipntu.github.io/belacon/). As part of this protocol, transcribers were instructed to segment and annotate all utterances during the parent-child interaction for the book reading, where an utterance is defined as any noise produced by the speaker by means of their vocal apparatus. The start and end of an utterance is defined by intonation patterns and pauses (Loh et al., 2022b). In the BELA transcription protocol (Loh et al., 2022a), a subdivision of an utterance, due to code switch to a different language, is known as a 'Grain'. Each utterance or subdivision is then labelled with boundaries for non-linguistic communicative acts including vocal sounds (e.g., humming), non-vocal sounds (e.g., clapping). Onsets and offsets of different languages are marked.

As the purpose of the transcription is for linguistic research within the Talk Together Study (Woon et al., 2021), recordings may sometimes contain regions that are neither transcribed nor labelled for languages spoken. In these regions, there are no ground-truth language labels or timestamps, and they will not be evaluated for the challenge.

Transcribers were given special instructions to place the start-stop boundaries carefully at the level of speaker turn, taking note to include sounds at the edge of words (such as fricatives at the end), as well as at the level of language information, taking note to include all word boundaries in each respective language in the event of language change (Figure 5). When overlap between speakers occurs, transcribers are instructed to ascertain the start and end of each speaker turn to the best of their abilities. In accordance with the BELA protocol (Loh et al., 2022a), each recording was transcribed by at least one transcriber and checked by a second transcriber. In the original transcription, all non-English utterances are translated to English (Figure 5). As machine translation and transcription falls outside the scope of the current Challenge, translations and transcriptions are omitted from the Challenge dataset.

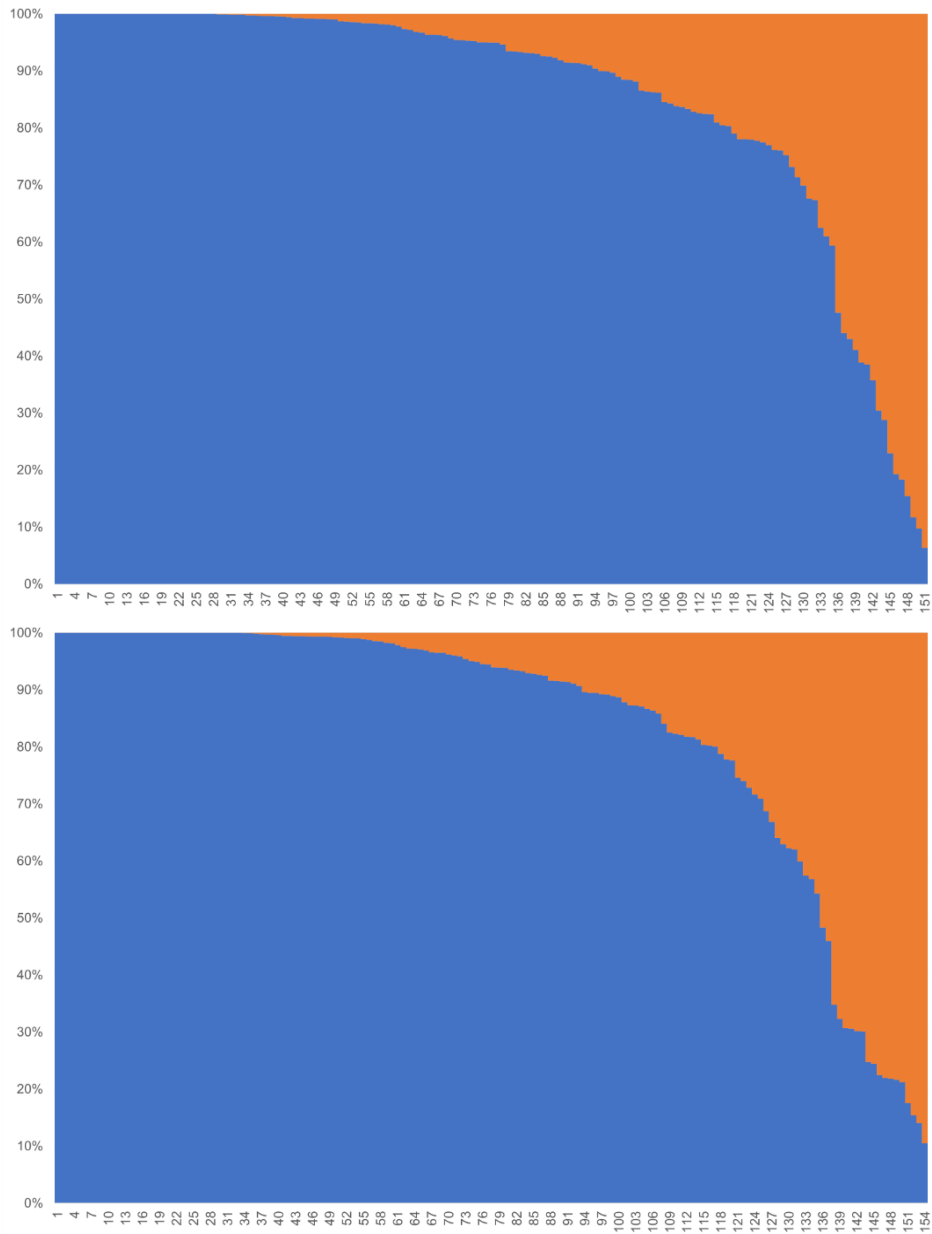

Figure 4. Proportion of English (blue) and Mandarin (orange) in each file, represented by each vertical bar, in the Development set (N = 151) [top] and Evaluation set (N = 154) [bottom].

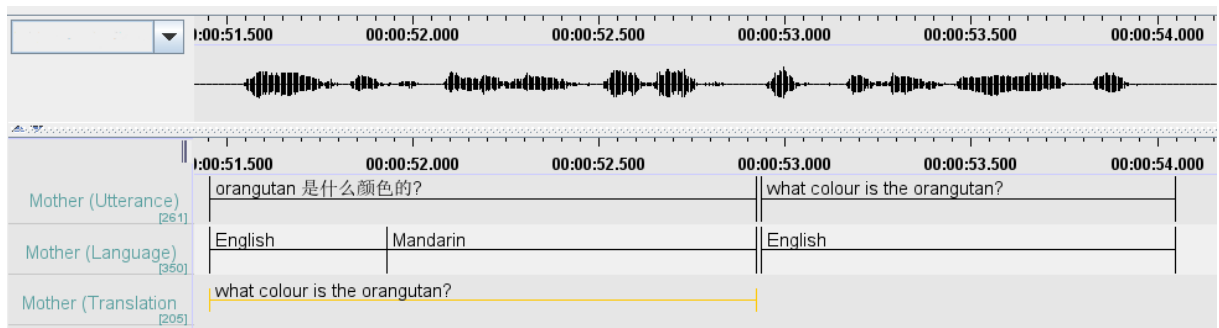

Figure 5. Transcription is done in ELAN. ELAN allows multiple tiers annotation within an utterance boundary. Note that for the purposes of the challenge, translations and transcriptions will not be released.

**Language Labels in the Challenge Datasets.** There are four language labels that will appear in the challenge dataset (Table 2). As English and Mandarin are target languages in the challenge, only regions in the recordings marked with these languages will be evaluated and scored in the challenge.

Table 2. Language labels that appear in the MERLIon challenge dataset. See following sections in the evaluation plan on how Languageless and Red-Dot annotation labels are recoded for the Challenge.

| English | Utterances labelled as English including Languageless and Red-Dot vocabulary, from the onset of English until the end of a prosodic unit of speech, or a switch to another language. *These labels are scored in the MERLIon Challenge.* |
|---|---|
| Mandarin | Utterances in Mandarin Chinese including Languageless and Red-Dot vocab, from the onset of English until the end of a prosodic unit of speech, or a switch to another language. *These labels are scored in the MERLIon Challenge* |
| Non-Evaluated-Speech | Utterances or their subdivisions originally labeled as Hokkien or Malay. *These labels are NOT SCORED in the MERLIon Challenge.* |
| Non-Speech | Utterances or their subdivisions originally labeled as Vocal Sounds (babbling, crying, laughter, etc.) or Non-Vocal Sounds (clapping, knocking, sounds made by an object). Non-Vocal sounds are only coded when they co-occur with an utterance, resulting in two shorter subdivisions. Utterance consisting only of redacted sensitive audio (replaced by a 440Hz sine wave tone) are also labeled Non-Speech. *These labels are NOT SCORED in the MERLIon Challenge.* |

The BELA transcription protocol included two unique language annotation types: "Languageless" vocabulary and "Red-Dot" vocabulary.

- Languageless – fillers and interjection words used by speakers regardless of language backgrounds (e.g., oh). a label for fillers such as "ah" and "oh" which may be used in any language were labelled as "languageless".

- Red Dot – for words used and understood commonly by local speakers in Singapore. These words may have originated from one of the local languages spoken in Singapore and is now widely used by local speakers regardless of language backgrounds (Woon & Styles, 2021).

  - Red Dot discourse markers (see full list in Loh et al., 2021c) are usually monosyllabic clause-final particles that convey mood, attitude, solidarity or emphasis (Gupta, 2006).

  - Red-Dot vocabulary (see full list in Loh et al., 2021c) includes words known by a large number of Singaporeans of all language backgrounds. Many Red-Dot vocab items are historic borrowings from regional varieties including Hokkien, Teochew, Cantonese, Malay. These words can be used when speaking any language, regardless of the language of origin, making them loans into each of the languages.

The Languageless and Red-Dot labels were designed to allow a variety of analyses relevant to linguistic structures in the local varieties but are not relevant to the task of language identification, as presented in the current Challenge. As such, the Challenge datasets have resolved these terms in the manner described below (Figure 6):

- Red-Dot discourse markers are recoded to the language label of the preceding word.

- In an utterance containing a mix of Languageless/Red Dot and target languages (English/Mandarin):

  - If the Languageless or Red Dot word is the first word, it is labeled with language of the segment after it.

  - If the Languageless or Red Dot word is the last word, it is labeled with the language of the segment preceding it.

  - If the Languageless or Red Dot word is between two same-language segments, it is labeled with that language.

  - If the Languageless or Red Dot word is between two different language segments, it is labeled with the language that has the longest duration in that utterance.

- In an utterance containing only Languageless/Red Dot, it is labeled with the main language used by the speaker in the entire file.

Figure 6. The Red Dot and Languageless labels (left) are recoded to English or Mandarin (right) based on the utterance it is in. Note that for the purposes of the challenge, translations and transcriptions will not be released.

**Personal Identifying Information.** The recordings in the Talk Together Study contain regions with personal identifying information (PII), which cannot be shared, to protect participant privacy in accordance with the ethical approval for the study (IRB-2018-10-001). In the challenge Development and Evaluation sets, the sensitive regions have been redacted by replacement with a sine wave tone at 440 Hz.

If an utterance consists of only a redacted speech segment, the utterance is labeled NON_SPEECH, and the utterance will not be scored. When redacted speech occurs within an utterance with language labels the following rule applies:

- If the redacted speech falls between two same-language segments, it is labeled with the surrounding language.
- If the redacted speech is between two different-language segments, it is labeled with the language that has the longest duration in that utterance.

For these utterances, the scoring regions consist of the entirety of the utterance minus the region of redaction. In both cases, the scoring regions will be specified and distributed as part of the Development and Evaluation releases.

**File formats and distributions.** All audio and relevant annotations (i.e., timestamps and language labels) of the MERLIon challenge will be distributed via the organizers. The datasets are hosted on Huawei Cloud. Upon registration in the challenge, the organizers will verify the registration details of each team and provide further instructions to the contact participant on each team via email.

The audio recordings are in single channel, WAV files, sampled at 32kHz, while reference annotations will be distributed as .csv files (Table 3). In the case of the Development set, reference timestamps and language labels will be provided. The start and end times of evaluated regions for Task 2 (Language Diarization) are also released. The evaluated regions are released in an Excel sheet (Table 4). The number of evaluated regions of each audio recording range from 1 to 2 (Table 4).

In the reference annotations for the Development set, there is a column (overlap_diff_lang) which indicates if audio segment overlaps with another audio segment with a different language label, and these audio segments are excluded from scoring for Task 1 (Language Identification). For instance, if an English segment overlaps with a Mandarin segment, both audio segments are tagged as True in the overlap_diff_lang column and will be excluded for calculation of equal error rate and balanced accuracy in the scoring of Task 1 (Language Identification) Development Set. These overlapping language segments are, however, included in the computation of language diarization error rate for Task 2 (Language Diarization).

In the Evaluation set, audio recordings and timestamps will be provided for Task 1 (Language Identification) and only audio recordings will be provided for Task 2 (Language Diarization). Note that the timestamp information given in the evaluation set for Task 1 (Language Identification) is unsuitable for Task 2 (Language Diarization). In the reference timestamps for Evaluation set of Task 1 (Language Identification), reference timestamps of audio segments that overlap with another audio segment with a different language label (e.g., English segment overlaps with a Mandarin segment) are not released at all. Reference timestamps of audio segments tagged as non-speech or non-evaluated languages will not be released as well. In the Evaluation set of Task 2 (Language Diarization), only audio recordings will be released. Audio segments that overlap with a different language label (e.g., English segment overlaps with a Mandarin segment) and non-speech segments tagged erroneously as English or Mandarin will be included in the computation of language diarization error rate and individual language error rates (see *Scoring*).

Table 3. Format of reference annotations given in development set.

| Audio_name | utt_id | Start time of annotation (ms) | End time of annotation (ms) | Language tag | Overlapping different language |
|------------|--------|-------------------------------|-----------------------------|--------------|--------------------------------|
| Audio1.wav | a1 | 1170 | 1500 | English | False |
| Audio1.wav | a2 | 1550 | 1780 | Mandarin | False |
| ... | ... | ... | ... | ... | ... |
| Audio2.wav | a1 | 0 | 900 | English | True |
| Audio2.wav | a2 | 800 | 2560 | Mandarin | True |
| ... | ... | ... | ... | ... | ... |
| Audio3.wav | a1 | 5 | 450 | Non-Speech | False |
| Audio3.wav | a2 | 1000 | 2560 | Mandarin | False |

Table 4. Format of evaluated regions for Task 2 (Language Diarization) given in development set. The first column denotes the audio recording filename and the second and third columns are the start and end times of each evaluated region respectively. Note that some audio recordings have more than 1 evaluated region.

| Audio1.wav | 1170 | 1500 |
|------------|------|------|
| Audio2.wav | 0 | 800 |
| Audio2.wav | 2500 | 490000 |
| Audio3.wav | 5 | 879000 |

**Baseline system**

The baseline system[5], provided by the MERLIon CCS organizers, is an end-to-end conformer model for both tasks. The model consists of four conformer encoder layers followed by a statistics pooling layer and three linear layers with ReLU activation in the first two linear layers. All self-attention encoder layers have eight attention heads with input and output dimensions being 512, and the inner layer of the position-wise feed-forward network is of dimensionality 2048. The 39-dimensional Mel Frequency Cepstral Coefficients (MFCC) features comprising 13-dim MFCCs and their first- and second-order deviations are extracted for each speech signal before being fed into the conformer encoder layers. The statistics pooling layer then generates a 1024-dimensional output which is finally projected by three linear layers to the number of target languages. The three linear layers comprise 1024, 512, and 2 output nodes.

The training data consists of the preselected partitions of 200 hours of Aishell Mandarin data, 100 hours of Librispeech data, and 100 hours of National Speech Corpus data as listed in the Training Data section of the evaluation plan. Prior to the feature extraction stage, speech signals in these datasets are segmented into maximum of 3s chunks. The model is trained for five epochs with batch size 32 and updated with a learning rate that warms up from 0 to 10^-4 in 5000 steps followed by the cosine annealing decay.

An energy-based voice activity detection is performed on the test data to identify the silent parts for the diarization task. Before performing language identification, each speech signal is partitioned into speech clips after removing silences. We assume there is no code-switch in each speech clip.

More information on the baseline system is hosted at the MERLIon CCS Github repository (https://github.com/MERLIon-Challenge/merlion-ccs-2023).

**Evaluation Rules**

Participation is open to all who are interested and comply with the evaluation rules as stated in the plan. Development and Evaluation data will be made available to registered participants, and the participants will process the data locally and submit the output of their systems for scoring. All participants must agree to process the data in accordance with the following rules:

- Participants agree to make at least one valid submission to the closed track of Task 1 (Language Identification) before the end of evaluation period. A valid result submission for Task 1 (Language Identification) is defined as one that contains prediction scores for all evaluated audio segments, passes the validation step during upload and receives a score.

---

[5] https://github.com/MERLIon-Challenge/merlion-ccs-2023/blob/master/readme.md#baseline-system

- In the closed tracks, participants are not allowed to use any additional data beyond the preselected training data outlined for system training or development.

- In the open tracks of both tasks, no more than 100 additional hours of training data can be used for system training or development.

- In the open tracks, if pretrained models are used, they must be publicly available and no more than 100 additional hours of data can be used for finetuning.

- Manual investigation of the Evaluation set (e.g., listening, segmentation or transcription), prior to the end of evaluation period is not allowed.

- Open-source code is strongly encouraged, but not required.

- Participants agree to follow the result submissions requirements. Results submissions will happen on CodaLab, and participants will need to create an account on CodaLab.

- During the evaluation period, each team may make at most 3 submissions per day.

- Participants agree to submit a system description document for all of their final systems. The document must clearly describe the training strategy (e.g., selection of data partitions), pretrained model name and version (if used), data processing, system architecture (e.g., algorithms and computational resources used), etc., in enough detail to be reproducible. These documents will be submitted at the end of the evaluation period.

For guidelines regarding the format of results files, refer to Appendix C and for the format of the system description, refer to Appendix D.

## Evaluation Protocol

Evaluation activities will be conducted over a CodaLab interface, which contains a result submission portal and leaderboard system. All files and folders must be formatted as instructed in *Scoring* and Appendix C. For additional instructions, please consult the MERLIon CCS website or reach out to us at merlion.challenge@gmail.com.

**Registration.** One participant from a team must register to participate in the evaluation, by providing the team's name, agreeing to the terms of participation, data use agreement and selecting the tasks they wish to participate in. Please note that the first person from the team that registers is deemed the team representative and data steward. As the team representative and data steward, the participant will be the main point of contact in the event of disputes and clarifications and ensure their team members comply with the terms set out in the data use and license agreement. Instructions for registration can be found at:

https://sites.google.com/view/merlion-ccs-challenge/registration?authuser=0

**Data use and license agreement.** One participant from each team must sign the data use and license agreement. After the data use and license agreement is verified by the organizers, the organizers will provide instructions for accessing the Development and Evaluation data. Teams must not share the access to the MERLIon CCS dataset to anyone not involved in the challenge.

**Results submission.** All result submissions will be conducted via CodaLab. Each team should only have one CodaLab account and all result submissions should be done via the account. More information and instructions on how to register an account and submit results via CodaLab is available on the MERLIon CCS website.

## Updates

Updates to this evaluation plan will be made available via the mailing list and the challenge website (https://sites.google.com/view/merlion-ccs-challenge).

## Appendix A

### Selection of partitions from the Aishell and Singapore National Speech Corpus and Download Instructions for SEAME corpus

The National Speech Corpus is headed by the Info-communications and Media Development Authority (IMDA) of Singapore, and it can be obtained via this link (https://www.imda.gov.sg/nationalspeechcorpus) on their website. Registration via filling up a form as well as a Dropbox account is required. The preselected partition for the MERLIon CCS Challenge is available at https://bit.ly/merlion-ccs-nsc-partition as well as https://web.hss.ntu.edu.sg/blip/MERLIon-CCS-Challenge_NSC-partition_v001.zip.

For the preselected partition, speakers were selected based on similar characteristics to the participants in the Talk Together Study (TTS) dataset of which the challenge development and evaluation datasets are derive from. Using an Excel spreadsheet, speakers of the National Speech Corpus were filtered based on their self-reported age, accent and occupation. To model the similar distribution of voices and speakers, participants whose age ranged from 20s to 40s with Chinese ethnicity were shortlisted. Mostly female participants were shortlisted, but some male voices were included too. Speakers who reported their employment status as students were removed. As the shortlisted speakers had more than 100 hours of audio data, the selection of the speakers for the 100 hours partitions was done by using RAND function on Microsoft Excel. Using the RAND function, each shortlisted speaker was assigned a random number. The random numbers assigned were frozen and arranged in ascending order. After which, speakers were picked one by one down the list the partition reached the targeted 100 hours.

For the Aishell corpus, participants were picked one by one down the list until the partition reached the targeted 200 hours. The filenames for the Aishell partition for the challenge is available at: https://bit.ly/merlion-ccs-aishell-filenames.

For the SEAME corpus, participants must register[6] with the MERLIon CCS challenge first. After registration, participants have to sign the LDC data license agreement[7] and send it back directly to LDC (ldc@ldc.upenn.edu) before gaining access. Note that the LDC data license agreement form is a separate form from the data license agreement form (for access to the MERLIon CCS Challenge Development and Evaluation sets.

---

[6] Registration to the MERLIon CCS Challenge can be done at: https://tinyurl.com/merlionccsregister
[7] LDC data license agreement for the SEAME corpus is available at https://bit.ly/merlionccs-seame-ldc-agreement

## Appendix B

### List of training corpora suggested

For the open tracks of the challenge, participants may use any publicly available or proprietary data to train their systems, as long as the additional amount of training data does not exceed 100 hours. If pretrained models are used, a maximum amount of 100 hours of data can be used for finetuning.

**English-Mandarin codeswitching corpora**

- ASCEND (https://huggingface.co/datasets/CAiRE/ASCEND)
- TALCS (https://arxiv.org/ftp/arxiv/papers/2206/2206.13135.pdf)

**Monolingual corpora**

*English*

- Singapore National Speech Corpus (https://www.imda.gov.sg/nationalspeechcorpus)
  - Please note that a preselected partition is used as training set for the closed tracks.
- LibriSpeech (http://www.openslr.org/12/)
  - Please note that a preselected partition is used as training set for the closed tracks.
- Speakers in the Wild (SITW) (http://www.speech.sri.com/projects/sitw/)
  - Please note that there is overlap between speakers in SITW and Voxceleb
- VoxCeleb (http://www.robots.ox.ac.uk/~vgg/data/voxceleb/)
- VoxCeleb 2 (http://www.robots.ox.ac.uk/~vgg/data/voxceleb/vox2.html)
- Common Voice (https://voice.mozilla.org/en/datasets)
- AVA ActiveSpeaker and AVASpeech (http://research.google.com/ava/)

*Mandarin*

- Aishell (https://www.openslr.org/33/)
  - Please note that a preselected partition is used as training set for the closed tracks.
- Chinese Corpus Set 1 (https://www.openslr.org/47/)
- aidatatang_200zh (https://www.openslr.org/62/)
- AliMeeting (https://www.openslr.org/119/)
- MAGICDATA Mandarin Chinese Conversational Speech Corpus (https://www.openslr.org/123/)

## Appendix C

## Results File Format Specification

For Task 1 (Language Identification), results for all audio segments to be evaluated should only be in a single .txt file named `prediction.txt`. All audio segments labelled as English and Mandarin and do not overlap with another language must be included. The order of the audio segments must be according to the order it is presented in the evaluation timestamps labels.

The audio segment id for each audio segment is the combination of audio filename and utt_id (as indicated in the utt_id column of the labels) and start and ends of that segment in milliseconds (as indicated in the start and ends columns). Taking the first audio segment in the MERLIon CCS Development set as an example:

| audio_name | utt_id | start | end |
|---|---|---|---|
| TTS_P91182TT_VCST_ECxxx_01_AO_48503281_v001_R004_CRR_MERLIon-CCS.wav | a1 | 1170 | 2750 |

The above audio segment should be named as:

**TTS_P91182TT_VCST_ECxxx_01_AO_48503281_v001_R004_CRR_MERLIon-CCS_a1_1170_2750**

There are two allowed formats for Task 1 results output. The first format requires each audio segment to have separate English and Mandarin scores per line. Each line has 3 space-delimited fields: audio segment id, followed by 0 (indicates English) or 1 (indicates Mandarin), and the respective language prediction score for each language. The English score (indicated by 0) must always precede the Mandarin score (indicated by 1). For instance:

```
TTS_P91182TT_VCST_ECxxx_01_AO_48503281_v001_R004_CRR_MERLIon-CCS_a1_1170_2750 0 4.21080
TTS_P91182TT_VCST_ECxxx_01_AO_48503281_v001_R004_CRR_MERLIon-CCS_a1_1170_2750 1 -10.018997
```

In the second allowed format for Task 1 (Language Identification), each line has 3 space-delimited fields, audio segment id followed by its English and Mandarin scores in the same line;

```
TTS_P91182TT_VCST_ECxxx_01_AO_48503281_v001_R004_CRR_MERLIon-CCS_a1_1170_2750 4.21080 -10.018997
```

Results for all audio segments to be evaluated should be enclosed in a single .txt file named `prediction.txt` and placed directly in a zip folder which should not contain any spaces in the folder name. The submission must have the following structure:

```
results.zip/

└── prediction.txt
```

For Task 2 (Language Diarization), a separate RTTM file should be generated for each audio recording. The RTTM file for each audio file should be named according to the audio filename. For instance, for the audio recording:

`TTS_P12345TT_VCST_ECxxx_01_AO_12345678_v001_R004_CRR_MERLIon-CCS.wav`

The corresponding RTTM file should be named:

`TTS_P12345TT_VCST_ECxxx_01_AO_12345678_v001_R004_CRR_MERLIon-CCS.txt`

In the RTTM file, each line contains three space-delimited fields, start time, end time and language id, indicating the onset and offset of language turns in milliseconds. For example:

```
413402.0 414611.0 English
415592.0 416762.0 Mandarin
417217.0 418350.0 English
```

All RTTM files to be evaluated must be placed in a zip folder (with no spaces in the filename) with the following structure:

`results.zip/`

`├── TTS_P12345TT_VCST_ECxxx_01_AO_12345678_v001_R004_CRR_MERLIon-CCS.txt`

`├── TTS_P22345TT_VCST_ECxxx_02_AO_45678910_v001_R007_CRR_MERLIon-CCS.txt`

`└── ...`

## Appendix D

## System Description Format

Each submitted system to the challenge must be accompanied by a system description which sufficiently details the system in a way for a fellow researcher to understand the approach and resources required to train and run the system.

If a team submits a system with largely similar components but specific tweaks for both Task 1 (Language Identification) and Task 2 (Language Diarization) respectively, it is fine to submit a system description document which describes the model architecture shared between both tasks under the respective subsections, detailing the specific modifications for each task and their performances in the respective tasks. For instance, if the only difference between the system architecture for Task 1 (Language Identification) and Task 2 (Language Diarization) is in the types of training data used, the system description should clearly outline that.

If a team submits two different systems for Task 1 (Language Identification) and Task 2 (Language Diarization) separately, it is fine to include both system descriptions in a single document. However, the document should clearly outline each system submitted. For instance, if the system architecture for Task 1 (Language Identification) is an E2E model while the system architecture in Task 2 (Language Diarization) is an adversarial learning model, both systems should have detailed descriptions of the training procedure and algorithm, clearly describing the different architecture structures.

If a team submits multiple systems for a task, not all systems must be included. However, the system description should include a comparison between the best performing systems submitted.

Deadline and instructions on system description submission will be updated on the MERLIon CCS website.

The system description should contain the following sections:

1. Authors – List of people whose contributions you wish to acknowledge.
2. Abstract – A few short sentences briefly describing the system. In this section, the tasks and tracks the system has been submitted to should be highlighted.
3. Brief summary of notable highlights – A short paragraph emphasising novel approaches or features used in the system that led to a significant improvement in performance. If multiple systems were submitted, differences amongst the systems submitted may be highlighted here.
4. Data resources

For open tracks: The data used for training including amounts and sources should be described. When possible, URLs and DOI links to datasets should be provided. If training data is selected from a larger corpus, selection processes of partition should be clearly detailed, i.e., it should be clear how the recordings were selected. If random selection or algorithmic selection has been performed, the procedure should be clearly described, e.g., first 100 hours of audio that fulfils a criterion or first 100 hours randomly selected.

In cases where a proprietary dataset is used for training or finetuning, it should be described in sufficient detail for the reader to get the gist of its composition in terms of speakers, speech registers and accents, languages spoken, recording environment and equipment, and speech elicitation protocol (e.g., reading versus conversational speech, types of topic prompts if any used).

If pre-trained models are used, the version and name of pretrained model should be provided, including a URL.

5. Detailed description of algorithm

Each system component should be described with sufficient information that another researcher would be able to reimplement it. If hyperparameter tuning was performed, a detailed description of the tuning process and final hyperparamters should be included.

For each major phase in the system, we suggest subsections:

     a. Voice activity detection
     b. Acoustic features – e.g., MFCCS, PLPs
     c. Signal processing – e.g., signal enhancement, denoising
     d. Segment representation – e.g., i-vectors
     e. Clustering method – e.g., agglomerative, k-means

6. Results on the development and evaluation sets

Teams must report performance of their systems on the MERLIon CCS challenge development and evaluation sets. For Task 1 (Language Identification), both equal error rate and balanced accuracy should be reported as output by the official scoring tool provided. For Task 2 (Language Diarization), language diarization error rate and the indiivdual language error rates for English and Mandarin should be provided as output by the official scoring tool provided.

Teams may also report results from additional systems. To encourage the development of robust systems, it is encouraged but optional to report results on additional datasets.